\documentclass[twocolumn,aps,prl,10pt,superscriptaddress,floatfix]{revtex4-1}
\usepackage{graphicx,subfigure}
\usepackage{amsmath,amssymb,bm}
\usepackage{mathtools}
\usepackage{multirow}
\usepackage{makecell}
\usepackage{textcomp}
\usepackage{float}
\usepackage{color}
\usepackage[normalem]{ulem}
\usepackage{dsfont}
\usepackage{mathrsfs}
\usepackage{braket}
\usepackage{transparent}
\usepackage{xcolor}
\usepackage{hhline}
\usepackage{graphicx}
\usepackage{pdfpages}
\usepackage[export]{adjustbox}
\usepackage{hyperref}
\usepackage{physics}

\makeatletter
\AtBeginDocument{\let\LS@rot\@undefined}
\makeatother

\usepackage{graphicx}

\graphicspath{ {figures/} }

\makeatletter
\let\saved@includegraphics\includegraphics
\AtBeginDocument{\let\includegraphics\saved@includegraphics}

\makeatother

\makeatletter
\newcommand*{\centerfloat}{%
  \parindent \z@
  \leftskip \z@ \@plus 1fil \@minus \textwidth
  \rightskip\leftskip
  \parfillskip \z@skip}
\makeatother

\usepackage{lineno} 
\begin{document}

\title{A Deep-Learning-Boosted Framework for Quantum Sensing with Nitrogen-Vacancy Centers in Diamond}

\author{
Changyu~Yao,$^{1,*}$
Haochen~Shen,$^{1,*}$
Zhongyuan~Liu,$^{1}$
Ruotian~Gong,$^{1}$
Md Shakil~Bin~Kashem,$^{1}$
Stella~Varnum,$^{2}$
Liangyu~Li,$^{3}$
Hangyue~Li,$^{3}$
Yue Yu,$^{1}$
Yizhou~Wang,$^{1}$
Xiaoshui~Lin,$^{1}$
Jonathan~Brestoff,$^{2,4}$
Chenyang~Lu,$^{3,5}$
Shankar~Mukherji,$^{1,4,6}$
Chuanwei~Zhang,$^{1,4}$
Chong~Zu$^{1,4,\dag}$
\\
\medskip
\normalsize{$^{1}$Department of Physics, Washington University, St. Louis, MO 63130, USA}\\
\normalsize{$^{2}$Department of Pathology and Immunology, Washington University School of Medicine, St. Louis, MO 63110, USA}\\
\normalsize{$^{3}$Computer Science \& Engineering, Washington University McKelvey School of Engineering, St. Louis, MO, USA, 63130}\\
\normalsize{$^{4}$Center for Quantum Leaps, Washington University, St. Louis, MO 63130, USA}\\
\normalsize{$^{5}$AI for Health Institute, Washington University, St. Louis, MO, USA., 63130}\\
\normalsize{$^{6}$Department of Cell Biology and Physiology, Washington University School of Medicine, St. Louis, MO 63110, USA}\\
\normalsize{$^*$These authors contribute equally to this work}\\
\normalsize{$^\dag$To whom correspondence should be addressed; E-mail: zu@wustl.edu}\\
}

\begin{abstract}
Nitrogen-vacancy (NV) centers in diamond are a versatile quantum sensing platform for high sensitivity measurements of magnetic fields, temperature and strain with nanoscale spatial resolution. 
A common bottleneck is the analysis of optically detected magnetic resonance (ODMR) spectra, where target quantities are encoded in resonance features. 
Conventional nonlinear fitting is often computationally expensive, sensitive to initialization, and prone to failure at low signal-to-noise ratio (SNR). 
Here we introduce a robust, efficient machine learning (ML) framework for real-time ODMR analysis based on a one-dimensional convolutional neural network (1D-CNN). The model performs direct parameter inference without initial guesses or iterative optimization, and is naturally parallelizable on graphics processing units (GPU) for high-throughput processing. 
We validate the approach on both synthetic and experimental datasets, showing improved throughput, accuracy and robustness than standard nonlinear fitting, with the largest gains in the low-SNR regime. 
We further validate our methods in two representative sensing applications: diagnosing intracellular temperature changes using nanodiamond probes and widefield magnetic imaging of superconducting vortices in a high-temperature superconductor. 
This deep-learning inference framework enables fast and reliable extraction of physical parameters from complex ODMR data and provides a scalable route to real-time quantum sensing and imaging.
\end{abstract}

\date{\today}

\maketitle

\section{Introduction}

Quantum sensing based on solid-state defects has emerged as a versatile platform for precision measurements, offering a unique combination of high sensitivity and nanoscale spatial resolution under ambient conditions \cite{Degen2017RMP,Wolfowicz2021NatRevMat,Doherty2013PhysRep,Schirhagl2014AnnuRev}.
Among these systems, the nitrogen-vacancy (NV) center in diamond has become a widely used workhorse because it can be optically initialized and read out and can maintain long-lived spin coherence at room temperature \cite{Doherty2013PhysRep,Rondin2014RepProgPhys,Barry2020RMP,Jelezko2006PSSA,Hanson2006PRB,Balasubramanian2009NatMater}.
The versatility of NV centers allows for the detection of diverse physical quantities, including magnetic fields, electric fields, temperature, and strain, with applications spanning condensed-matter physics and life sciences, from imaging magnetic textures and superconducting phenomena to monitoring intracellular thermodynamics \cite{Taylor2008NatPhys,Maze2008Nature,Dolde2011NatPhys,Kucsko2013Nature,Ovartchaiyapong2014NatComm,Hsieh2019Science,Dovzhenko2018NatComm,Jenkins2019PRMat,Thiel2016NatNano,LeSage2013Nature,Glenn2015NatMethods,Hsieh2019Science}.

\begin{figure*}
    \centering
    \includegraphics[width=0.75\textwidth]{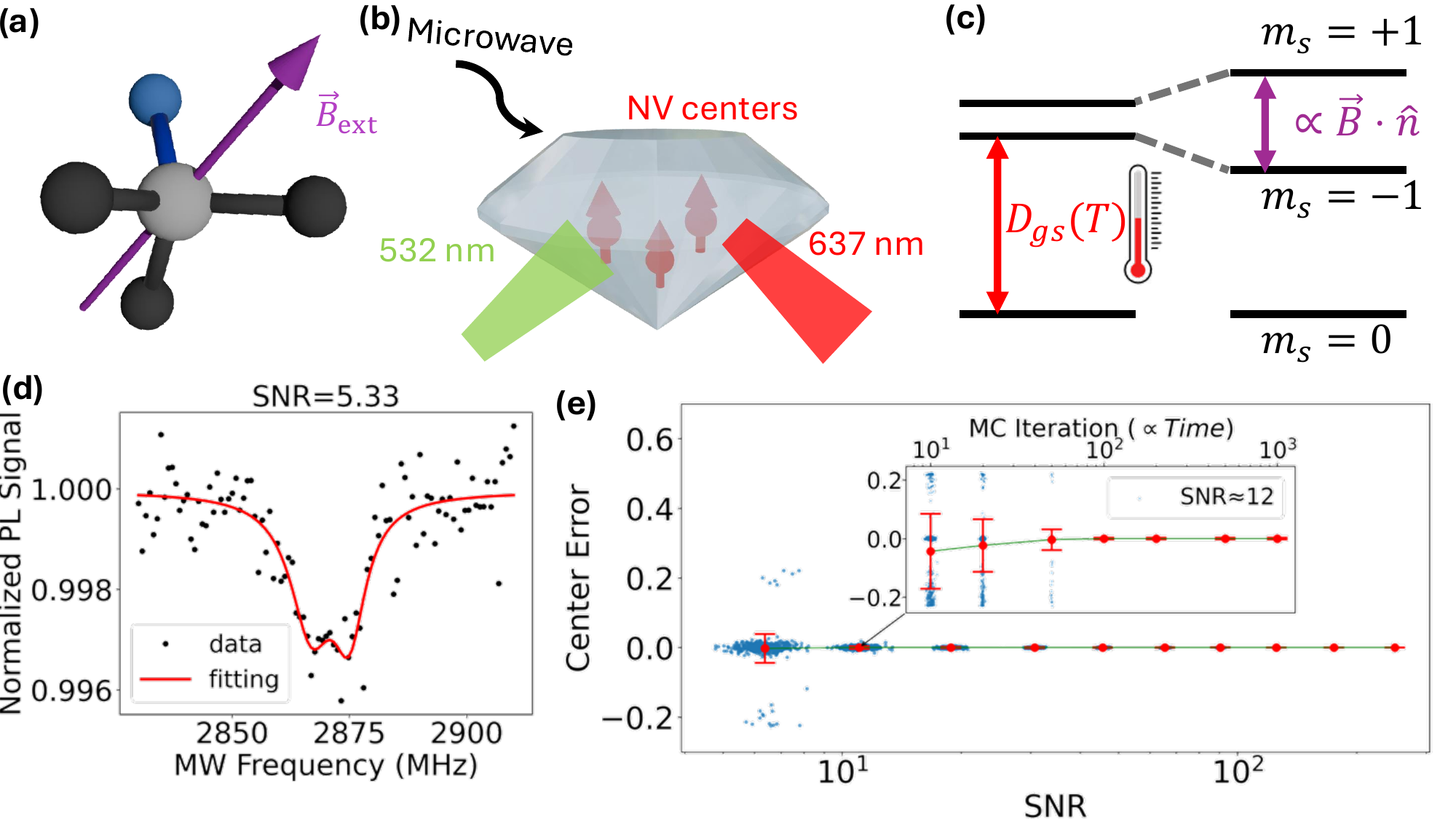}
    \caption{{\bf NV-Center Sensing Scheme and Monte-Carlo Fitting Characterization}
    (a) An NV center in a diamond lattice. An NV center is a substitutional nitrogen (N) atom adjacent to a lattice vacancy (V). The lattice structure illustrates the four possible NV crystallographic orientations.
    (b) Schematic of NV centers in diamond under optical excitation. A 532~nm laser initializes and reads out the spin state via spin-dependent photoluminescence (PL), while microwave (MW) fields drive spin transitions.
    (c) Ground-state energy level structure of the NV center. 
    The spin-triplet ground state ($S=1$) exhibits a zero-field splitting $D_{gs} \approx 2.87$~GHz at room temperature and the splitting is sensitive to temperature with a typical coefficient $\mathrm{d}D_{gs}/\mathrm{d}T \approx -70$~kHz/K. 
    Furthermore, the $|m_s=\pm1\rangle$ states degeneracy is lifted via the Zeeman effect with splitting $\delta = 2 \gamma_{e} B_{\parallel}$ under external magnetic field.
    (d) Representative ODMR spectrum of NV ensemble in a nanodiamond displaying the characteristic double-dip feature of $|0\rangle \leftrightarrow |\pm 1\rangle$ transitions. Here we intentionally choose a spectrum with low SNR, $\mathrm{SNR}=5.33$
    (e) Estimation error as a function of SNR (fixed at 200 Monte Carlo (MC) iterations).
    \textbf{Inset:} Convergence of MC fitting error versus iteration count at fixed SNR ($\approx 12$), showing reduced variance, at the expense of increased computational latency.
    \label{fig:NV}}
\end{figure*}

Many NV sensing experiments rely on continuous-wave optically detected magnetic resonance (ODMR), where shifts and splittings of fluorescence features encode the underlying physical parameters \cite{Rondin2014RepProgPhys,Barry2020RMP,ElElla2017OptEx,Jensen2013PRB}.
Extracting these quantities is typically performed by fitting ODMR spectra with phenomenological line shapes, such as sums of Lorentzian or Gaussian functions, using nonlinear least-squares methods \cite{Jensen2013PRB,Sengottuvel2022SciRep}.
Although robust for single-point, high signal-to-noise ratio (SNR) measurements, this workflow can become a practical limitation as data volumes grow, for example in widefield imaging or time-resolved experiments that require repeated analysis across many pixels and time points \cite{LeSage2013Nature,Glenn2015NatMethods,Wojciechowski2018RSI,Parashar2022SciRep}.
In these regimes, iterative fitting increases computational overhead and often demands careful initialization and parameter constraints to avoid failures or biased solutions, especially when spectra are degraded by noise and experimental drifts (Fig.~\ref{fig:NV}) \cite{Sengottuvel2022SciRep,Wojciechowski2018RSI}.

Machine-learning (ML) approaches provide an alternative route to spectral inference by learning a direct mapping from measured spectra to target parameters after an offline training stage \cite{Homrighausen2023Sensors,Tsukamoto2022SciRepML,Zhang2024PRA,Santagati2019PRX,Dushenko2020PRApplied,Rajpal2025OptExpress, wang2017time, yamamoto2025nanodiamond}.
Once trained, neural networks can evaluate spectra with a fixed computational cost per input and can be engineered to be tolerant to common noise sources through appropriate training data and regularization, bypassing the need for iterative optimization \cite{Homrighausen2023Sensors,Tsukamoto2022SciRepML,Zhang2024PRA,Qian2021APL}.

\begin{figure*}
    \centering
    \includegraphics[width=0.8\textwidth]{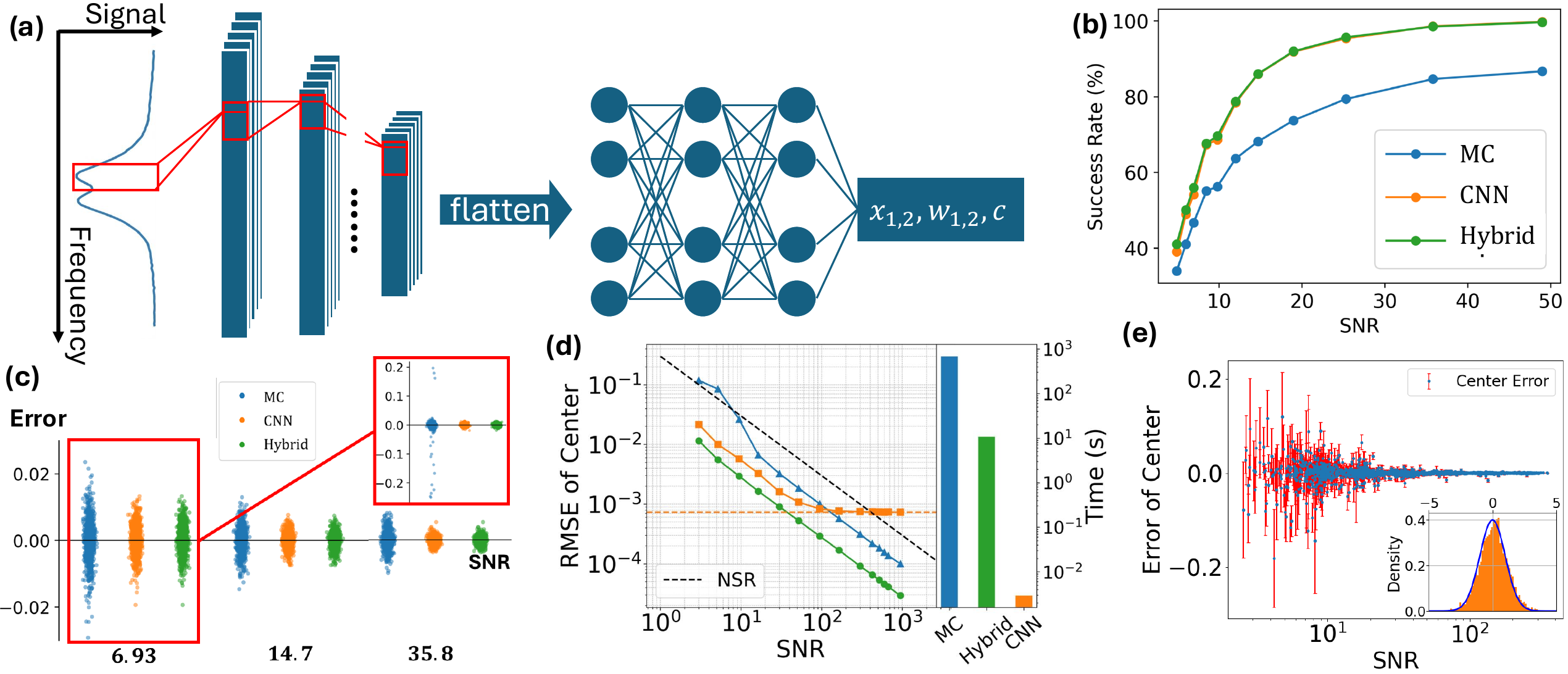}
    \caption{{\bf Model Architecture and Performance Validation on Synthetic Data}
    (a) Schematic of the proposed 1D-CNN architecture, comprising 5 convolutional layers for feature extraction followed by 3 fully connected layers ($\sim 70$ M parameters).
    (b) Prediction success rate for center frequency as a function of SNR ($N=1,024$). A prediction is classified as successful if the absolute center frequency error is $< 0.003$ (normalized frequency range to $[0,1]$).
    (c) Scatter plot of estimation errors for the center frequency across varying SNR levels.
    \textbf{Inset:} Expanded view of the error distribution at SNR $= 6.93$, highlighting significant outliers attributed to Monte-Carlo fitting failures.
    (d) RMSE of the estimated center frequency versus SNR. The black dashed line indicates the expected Poisson noise scaling, while colored dashed lines represent the algorithmic error floor determined from noise-free synthetic data, due to model capacity or floating point error.
    \textbf{Inset:} Wall-clock runtime comparison between Monte Carlo (MC) fitting (683~s), CNN inference (2.94~ms), and the Hybrid approach (10.8~s) for 5000 spectra.
    (e) Prediction error of the center frequency versus SNR. Red error bars represent the CNN-predicted aleatoric uncertainty ($1\sigma$) for individual spectra, illustrating the expected decrease in uncertainty at higher signal levels.
    \textbf{Inset:} Distribution of standardized residuals. The close agreement with a standard normal distribution validates the calibration of the model's aleatoric uncertainty estimates.
    \label{fig:CNN}}
\end{figure*}

Motivated by these opportunities, here we develop a one-dimensional convolutional neural network (1D-CNN) for ODMR analysis and benchmark its performance against standard fitting procedures across regimes relevant to NV sensing.
Our main results are in three folds. First, we develop a 1D-CNN architecture for ODMR analysis (Fig~\ref{fig:CNN}) and benchmark its performance on synthetic datasets across a range of SNR conditions, comparing prediction accuracy with conventional nonlinear least-squares fitting. These tests quantify improvements in analysis throughput and highlight enhanced precision and stability, particularly in low-SNR regimes.
Second, we validate the approach on experimentally measured ODMR spectra from NV centers in nanodiamonds, demonstrating reliable performance on real data and improved resilience to noise relative to least-squares fitting (Fig.~\ref{fig:T}a).
Finally, we showcase the breadth of the framework in representative quantum sensing scenarios: nanodiamond-based thermometry, spanning large-scale sensor calibration and intracellular temperature measurements in mouse macrophages (Fig.~\ref{fig:T})~\cite{kashem2025multiplexed}, and widefield magnetometry for imaging superconducting vortices in a high-temperature superconductor (Fig.~\ref{fig:B})~\cite{liu2025quantum}.

\section{Nitrogen-Vacancy Centers in Diamond}

The nitrogen-vacancy (NV) center in diamond is a point defect formed by a substitutional nitrogen atom adjacent to a lattice vacancy (Fig.~\ref{fig:NV}a)~\cite{Doherty2013PhysRep}.
The defect hosts a spin-triplet ground state ($S=1$) with a zero-field splitting (ZFS) of $D \approx 2.87$~GHz between the $|m_s=0\rangle$ and $|m_s=\pm1\rangle$ sublevels (Fig.~\ref{fig:NV}c). 
Under 532~nm excitation, a spin-dependent intersystem crossing enables efficient optical polarization into the $|m_s=0\rangle$ state and readout through spin-dependent photoluminescence, where $|m_s=0\rangle$ emits more photons on average than $|m_s=\pm1\rangle$.
The spin transitions can be measured using optically detected magnetic resonance (ODMR): by sweeping the frequency of an applied microwave field while monitoring the fluorescence intensity, a drop in fluorescence is observed when the microwave is resonant with the $|m_s=0\rangle \leftrightarrow |m_s=\pm1\rangle$ transitions. 
An external magnetic field lifts the degeneracy of the $|m_s=\pm1\rangle$ states via Zeeman splitting (proportional to the magnetic field on the NV axis), while temperature induces a common-mode shift of the ZFS. These responses allow the NV center to simultaneously probe the local magnetic field and temperature environment.
Detailed Hamiltonian is further discussed in the Supplementary Materials. 

Figure~\ref{fig:NV}d shows a representative ODMR spectrum taken from an ensemble of NV centers in a nanodiamond in the absence of external magnetic field.
In this case, the characteristic feature of the ODMR spectrum displays two largely overlaped resonance peaks with a small splitting originated from the local electric field environment from nearby charged defects inside the diamond.
The signal quality is inherently limited by photon shot noise, which scales with the square root of the collected fluorescence count rate, $~\sqrt{N}$.
To quantify the quality of ODMR spectra, we therefore define the SNR as $\text{SNR} = C \sqrt{N}$, where $C$ denotes the optical readout contrast of the ODMR resonance.
Here, we intentionally present a low-SNR spectrum ($\mathrm{SNR}=5.33$) to emulate realistic experimental conditions, such as intracellular nanodiamond sensing, where only limited averaging can be performed.

Conventional analysis to extract resonance frequencies typically relies on nonlinear least-squares fitting of ODMR spectra to a phenomenological line shape.
In this approach, one minimizes the mean-squared error (MSE) between the measured data and a model function, here taken as a sum of two Lorentzian peaks.
However, this iterative procedure is sensitive to the choice of initial parameter guesses and can converge to local minima, especially in low-SNR conditions where photon shot noise obscures the resonance features.

To mitigate sensitivity to the initial parameter guesses, a Monte Carlo (MC) strategy is often employed, in which the fitting routine is repeated with many randomized initializations and the solution yielding the lowest MSE is selected.
This added robustness comes at the cost of substantial computational overhead, as the runtime scales approximately linearly with the number of MC iterations.
As illustrated in Fig.~\ref{fig:NV}e, the relative fitting error at a fixed 200 MC iterations increases sharply in the low-SNR regime.
Reducing this error generally requires increasing the number of MC rounds, which in turn increases the computation time (Fig.~\ref{fig:NV}e, inset).
As a trade-off between accuracy and efficiency, we adopt 200 MC rounds as the baseline for benchmarking conventional fitting throughout this study.

\section{1D Convolutional Neural Network Framework}

In this work, we employ a one-dimensional convolutional neural network (1D-CNN) as the core architecture for ODMR spectrum analysis.
This choice is motivated by the structure of ODMR spectra, where key physical parameters (resonance frequencies and linewidths) are encoded in localized, approximately translation-invariant features (ODMR spectral dips) \cite{Fukushima1980Neocognitron,LeCun1998ProcIEEE}.
CNNs are well suited to capture such local geometric patterns regardless of their position within the sampled window \cite{LeCun1998ProcIEEE}.
The detailed network architecture is illustrated in Fig.~\ref{fig:CNN}(a).
It consists of five convolutional layers followed by three fully connected (FC) layers, with a total of $\sim 70$ million trainable parameters.
We performed a comparative analysis on the network depth: while a 3-layer architecture struggles to resolve the complexity of overlapping peaks, a 4-layer design yields acceptable performance.
Increasing to five layers provides a modest but consistent improvement in prediction accuracy (see Supplementary Information); because the added computational overhead remained manageable, we adopt the 5-layer design to maximize robustness.
Kernel sizes (ranging from 7 to 13) are chosen to span the maximum expected linewidth ($<10$~\% of the entire microwave sweeping range) while avoiding excessive sizes that might degrade local positional information.
Consequently, the precise retrieval of the resonance frequency relies heavily on the FC layers to map the extracted features to global coordinates.
This architecture implicitly emphasizes spectral features near the center of the microwave frequency sampling window, which matches a typical experimental ODMR protocol where the microwave sweep is centered around the expected resonance frequency.
In the Supplementary Materials, we list the detailed parameters used and do a thorough study on model parameters and architectures.

Each input spectrum comprises 101 uniformly sampled microwave frequency points, with the frequency axis normalized to $[0,1]$.
Before being passed to the network, the raw fluorescence signal $I$ is Z-score normalized as $I_{\mathrm{norm}}=(I-\mu)/\sigma$.
This preprocessing suppresses amplitude variations, expands the effective dynamic range, and encourages the model to focus on spectral line shape rather than absolute intensity.
Although Z-score normalization removes the absolute contrast $C$ in ODMR, it substantially improves robustness to variations in microwave delivery power and fluorescence collection efficiency, which are common in practical measurements.
To reduce overfitting and improve generalization to experimental data, we train the model using an on-the-fly synthetic data pipeline \cite{Srivastava2014Dropout,Tobin2017DomainRandomization}.
Rather than relying on a fixed pre-generated dataset, synthetic spectra are generated dynamically during training with randomized parameters within physically relevant ranges: normalized center frequency $\in[0.35,0.65]$, splitting $\delta\in[0.02,0.2]$, linewidth $w\in[0.02,0.09]$, and contrast $C\in[0.012,0.15]$.
To enforce realistic double-peak ODMR structures, we impose constraints such that the splitting is at least $0.8\times$ the average linewidth, and the width/contrast ratios between the two peaks are maintained above $0.75$  (see Supplementary Materials for details).
The model is trained for 300 epochs with batch size 128 and 256 batches per epoch, requiring approximately 21 hours on a single NVIDIA GeForce RTX 4070 GPU.

\section{Performance validation on synthetic data}

In this section, we compare three strategies for extracting physical parameters from ODMR spectra: conventional MC fitting, direct neural-network inference, and a hybrid approach that combines the two.
In the hybrid scheme, the 1D-CNN prediction serves as the initial guess parameters for the fitting, thereby coupling the CNN model’s global inference capability with the local precision of iterative optimization.

We first validate the CNN model using synthetic double-Lorentzian spectra spanning a wide range of SNRs.
Figure~\ref{fig:CNN}b summarizes the prediction success rate for 1,024 test spectra.
A prediction is considered successful when the absolute error in the center frequency is below $0.003$ (normalized units).
For a typical sweep range of $2.8$--$2.92$~GHz, this criterion corresponds to a temperature resolution of approximately $5$~K, which serves as a practically relevant benchmark for biological thermometry.
Although all three methods degrade at low SNR, the ML-based approaches (both CNN and hybrid) maintain a clear advantage over MC fitting.
We attribute this improvement to the CNN’s ability to extract global spectral features, which reduces susceptibility to the local-minimum traps that often hinder gradient-based fitting algorithm at moderate noise levels.
At higher SNR regimes, while the success rates of all methods will eventually converge to unity ($100\%$), the ML-based approaches reach this asymptote more rapidly.
This performance advantage is further corroborated by the error distributions shown in Fig.~\ref{fig:CNN}c, where ML-based methods display a narrower spread across all SNR levels.
Notably, the inset of Fig.~\ref{fig:CNN}c reveals that MC fitting is prone to catastrophic outliers, particularly at low SNR regimes, consistent with occasional convergence to spurious local minima.

We further quantify the performance by comparing the root-mean-square error (RMSE) to the Poisson noise-limit scaling (Fig.~\ref{fig:CNN}d).
At low SNR, all three approaches are fundamentally noise-limited and exhibit the same scaling behavior $\sim \text{SNR}^{-1}$.
At high SNR, however, the standalone CNN approaches a systematic error floor, consistent with residual regression bias, which we attribute to the finite spectral sampling resolution and limited model capacity.
The hybrid approach mitigates this limitation by utilizing the CNN output as the initialization for nonlinear fitting, thereby recovering fitting-level precision in the high-SNR regime.

Beyond ultimate accuracy, a key advantage of the neural network framework is that parameter extraction is decoupled from iterative optimization.
In traditional fitting method, the nonlinear optimizer must be executed repeatedly, so the computational cost scales unfavorably with both the number of spectra and the number of MC initializations.
In contrast, direct inference maps each spectrum to its parameters in a single forward pass.
This enables fully parallelized execution on GPUs, reducing the time complexity by several orders of magnitude (Fig.~\ref{fig:CNN}d, inset).
Specifically, processing 5000 spectra using MC fitting (200 iterations) with 8 multiprocessing workers requires 683~s.
In comparison, pure 1D-CNN inference processes the same 5,000 spectra in just 2.94~ms, while the hybrid method, which refines the CNN predictions via nonlinear fitting, completes the task in 10.8~s with no multiprocessing.
This massive acceleration makes real-time analysis of large-scale widefield quantum imaging data practical.

In addition, by incorporating a probabilistic loss function during training, the model provides per-spectrum uncertainty estimates.
As expected, the predicted uncertainty decreases systematically with increasing SNR (Fig.~\ref{fig:CNN}e), reflecting improved estimation precision with longer data acquisition times.
The reliability of these confidence intervals is validated through the distribution of standardized errors, defined as $(\varepsilon_{\text{pred}} - \overline{\varepsilon}) / \sigma_{\text{pred}}$.
The resulting distribution closely follows a standard normal profile (Fig.~\ref{fig:CNN}e, inset), indicating that the predicted uncertainties accurately capture the aleatoric noise in the parameter estimates.
Collectively, these results show that our ML-based framework not only dramatically accelerates ODMR analysis, but also delivers statistically rigorous uncertainty estimates that are essential for robust physical measurements.

\begin{figure*}
    \centering
    \includegraphics[width=0.94\textwidth]{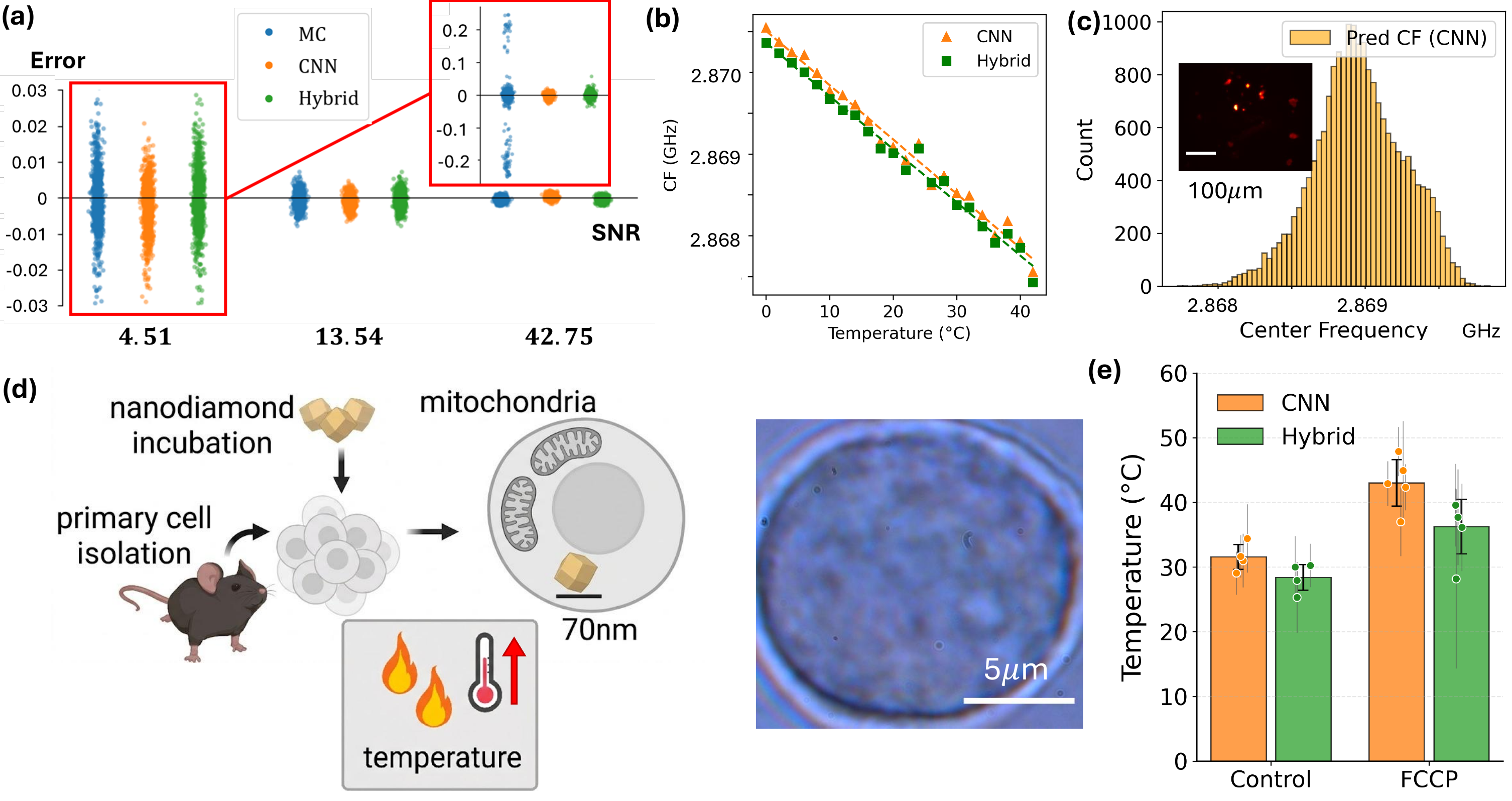}
    \caption{{\bf Calibration on Experimental Data and \textit{In Vivo} Thermometry}
    (a) Scatter plot of estimation errors across varying SNR levels. The high-SNR reference dataset (SNR $\approx 249$) serves as the ground truth.
    \textbf{Inset:} Expanded view of the error distribution at SNR $= 4.51$, highlighting significant outliers attributed to Monte-Carlo fitting failures.
    (b) Temperature dependence of the center frequency (Zero-Field Splitting, $D$). Linear regression of the Machine Learning inference (orange) and hybrid fitting (green) yields thermal coefficients $dD/dT = -66.5$ kHz/$^\circ$C ($R^2 = 0.9844$) and $-65.0$ kHz/$^\circ$C ($R^2 = 0.9859$), respectively.
    (c) Histogram of pixel-wise center frequencies at $T = 32^\circ$C ($N=19,618$ pixels, SNR $>5$). The distribution is centered at $2.8689$ GHz with a Full-Width at Half-Maximum (FWHM) around $0.5$ MHz, corresponding to a temperature variance of $\pm3.8$ $^\circ$C.
    \textbf{Inset:} Wide-field fluorescence image of drop-cast nanodiamond clusters, averaged over 2,000 frames.
    (d) \textbf{Validation in biological systems.} Schematic of ND bio-sensing and FCCP-induced thermogenesis. FCCP dissipates the mitochondrial proton gradient, thereby releasing metabolic energy as heat. Consequently, FCCP-treated cells exhibit elevated intracellular temperatures compared to wild-type controls.
    \textbf{Right panel:} Image of peritoneal exudate cell under white light.
    (e) Intracellular temperature measurements comparing $4$ untreated control and $5$ FCCP-treated cells. Hybrid fitting extracts average temperatures of $28.4 \pm 2.3^\circ$C (untreated control) and $36.2 \pm 4.7^\circ$C (FCCP). In comparison, the CNN prediction yields $31.8 \pm 2.2^\circ$C (untreated control) and $43.8 \pm 4.3^\circ$C (FCCP). Both methods successfully capture the relative warming trend induced by the uncoupler. P-values between untreated control and FCCP-treated groups: CNN: 0.0013, Hybrid: 0.017. P-values between methods: untreated control: 0.21, FCCP: 0.14.
    \label{fig:T}}
\end{figure*}

\section{Experimental Validation and Quantum Sensing Applications}

Having established the accuracy, robustness, and computational efficiency of our neural-network framework using synthetic data, we next assess its performance on experimental ODMR spectra in practical NV sensing scenarios.
In this section, we apply the pre-trained 1D-CNN directly to experimentally measured ODMR spectra without any empirical fine-tuning.
We first benchmark the model's estimation accuracy over a range of experimental SNR to evaluate its robustness to real-world noise.
We then demonstrate the framework’s versatility and throughput in two representative sensing modalities: characterizing intracellular temperature changes in mouse macrophages and spatial mapping of magnetic field distributions in superconductors.

We start by comparing 1D-CNN's performance against traditional fitting methods using a well-characterized experimental dataset.
Specifically, we acquire a high-quality reference ODMR spectrum with extensive signal averaging using an ensemble of NV centers in a nanodiamond, yielding an exceptional high $\text{SNR}\approx 249$.
Parameters extracted from this near-noiseless spectrum serve as our experimental ground truth.
By evaluating spectra with progressively shorter integration times thus lower SNR, we quantify how the estimation error scales under realistic measurement conditions.
Figure~\ref{fig:T}a shows the scatter plot of these estimation errors as a function of SNR.
Consistent with the synthetic benchmarks, the ML-based approach exhibits improved stability as the signal quality degrades.
In the low-SNR regime (SNR $=4.51$), the contrast between the methods becomes particularly pronounced, as highlighted by the expanded error distribution in Fig.~\ref{fig:T}a inset.

\emph{Nanodiamond Thermometry for Biology}---We next apply NV centers in nanodiamonds as nanoscale thermometers for biological sensing. Nanodiamonds are well suited for bio-quantum sensing owing to their chemical stability, biocompatibility, and ease of cellular internalization.
To establish a reliable baseline, we first calibrate the thermal response of a small cluster of 70~nm nanodiamonds drop-cast onto a glass substrate.
The substrate temperature is increased in small discrete steps, allowing sufficient time for thermal equilibration at each set point before ODMR acquisition.
The 1D-CNN extracts the temperature-dependent ZFS, $D(T)$ (Fig.~\ref{fig:T}b), yielding a thermal susceptibility of $dD(T)/dT \approx -66.5$~kHz/$^\circ$C near room temperature.
This value agrees well with the hybrid-fitting result ($-65.0$~kHz/$^\circ$C) and is consistent with typical literature reports ($\approx -70$~kHz/$^\circ$C).

Owing to variations in the local stress environment, individual nanodiamonds can exhibit slightly different intrinsic properties, leading to particle-to-particle inhomogeneity in the ZFS.~\cite{foy2020wide}
To quantify the intrinsic ZFS distribution of the nanodiamond ensemble, we performed widefield ODMR imaging at a fixed temperature of $32^\circ$C.
This dataset comprises ODMR spectra from $19,618$ pixels with SNR $> 5$, making it particularly well suited for validating our CNN approach under large-scale, high-throughput inference.
Figure~\ref{fig:T}c shows a histogram of the extracted ZFS, which is well described by a Gaussian distribution.
The full width at half maximum (FWHM) is $\sim 0.5$~MHz, corresponding to an apparent temperature variance $\pm 3.8 ^\circ$C.
This broadening is dominated by variations in local lattice strain and defect environment across individual nanodiamonds.
We therefore treat this inhomogeneous broadening ($\pm 3.8 ^\circ$C) as the systematic uncertainty floor for ensemble thermometry.

Despite this substantial intrinsic variance, averaging across multiple nanodiamonds internalized within a living cell still enables reliable detection of cellular temperature upon the activation of key biological processes.
To validate our CNN model, we analyze comparative thermometry ODMR dataset on two groups of wildtype (WT) mouse peritoneal macrophages that engulfed nanodiamond sensors: an untreated control group and a treated group exposed to carbonyl cyanide-p-trifluoromethoxyphenylhydrazone (FCCP)~\cite{kashem2025multiplexed}.
FCCP is a potent mitochondrial protonophore that dissipates the proton gradient across the inner mitochondrial membrane, thereby uncoupling the electron transport chain from ATP synthesis (Fig.~\ref{fig:T}d).
This uncoupling diverts metabolic energy from ATP production to heat generation, leading to a net increase in intracellular temperature \cite{Luvisetto1987Biochemistry_FCCPUncoupling,Tsuji2017SciRep_FCCPIntracellularHeating,Consiglio2026ACSApplNanoMater_ODMR_FCCP_Thermometry}.
Figure~\ref{fig:T}e summarizes the \textit{in vivo} thermometry results.
To approximate the global cellular thermal state and reduce the impact of particle-to-particle heterogeneity, each data point represents the mean apparent temperature of a single cell, obtained by probing $4$--$5$ distinct nanodiamond clusters distributed throughout the cell volume.
Hybrid methods yields $28.4 \pm 2.3~^\circ$C for the untreated control group and $36.2 \pm 4.7~^\circ$C for the FCCP-treated group. We consider these hybrid results to be the most accurate benchmarks among traditional fitting techniques.
The 1D-CNN revealed a similar temperature elevation, yielding mean extracted temperatures of $31.8 \pm 2.2~^\circ$C (untreated control) and $43.8 \pm 4.3~^\circ$C (FCCP).
These results indicate that, even under substantial systematic variance and complex experimental noise, the proposed ML-based framework can reliably recover underlying thermodynamic trends.

\begin{figure*}
    \centering
    \includegraphics[width=0.9\textwidth]{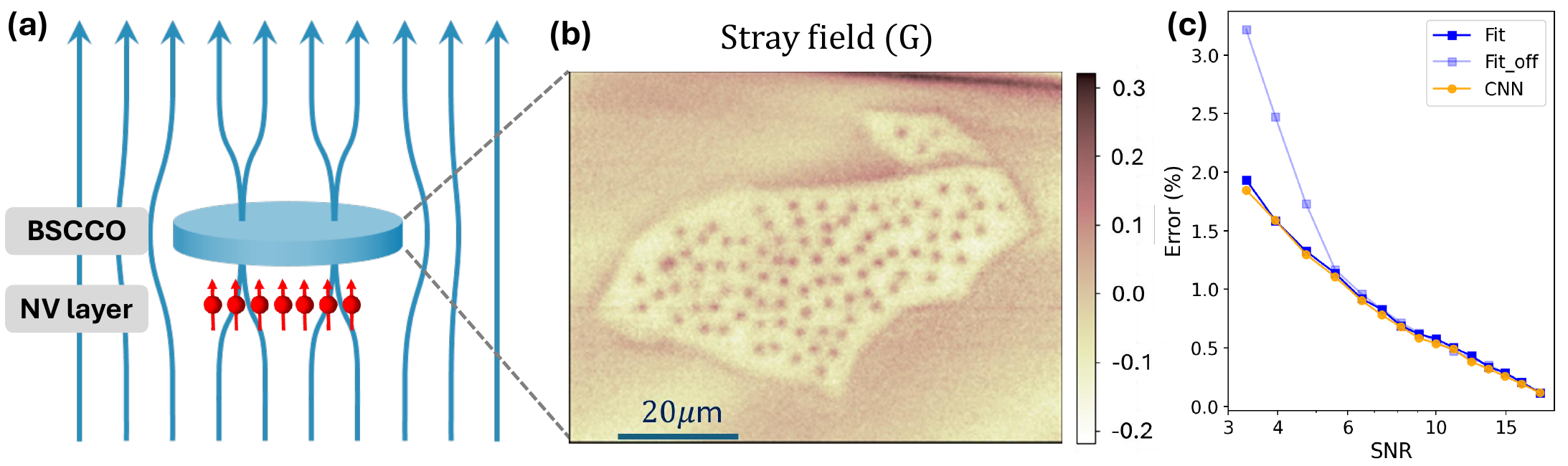}
    \caption{{\bf Magnetic Field Imaging of Superconducting Vortices using NV Centers}
    (a) Illustration of the Meissner effect and vortex formation in a Type-II superconductor. External magnetic field lines are expelled from the bulk material, while discrete magnetic flux quanta penetrate the superconductor in the mixed state.
    (b) Magnetic stray field map of the BSCCO flake reconstructed by the 1D-CNN at $T=60$ K under a $1.25$ G bias field. Superconducting vortices are clearly resolved as distinct dark dots.
    (c) Reconstruction error (RMSE) of the magnetic field map as a function of SNR. The network inference (orange) is compared with single-shot least-squares fitting using optimized (dark blue, center $= 0.5$) and perturbed (light blue, center $= 0.45$) initial guesses. While the network maintains low error across the range, the fitting method with poor initialization degrades significantly at low SNR. SNR levels correspond to different signal averaging times, with the dataset from the longest integration time serving as the ground truth reference.
    \label{fig:B}}
\end{figure*}

\emph{Widefield Imaging of Superconducting Vortices}—We next consider a second quantum sensing scenario: magnetometry using a thin layer of near surface NV ensembles in a diamond plate.
Widefield NV magnetometry has emerged as a powerful tool for studying condensed matter systems, from quantum magnetism to superconductivity, by enabling \textit{in situ} imaging of sample inhomogeneity and local magnetic textures with sub-micron spatial resolution.
Moreover, integrating NV centers into diamond anvil cells provides a new lens to \textbf{locally} characterize quantum materials under extreme pressures, accessing regimes beyond thermally driven transitions and into pressure-tuned phenomena.

In widefield NV magnetometry, the fluorescence from a large field of view of NV centers is imaged onto a camera, producing a high-dimensional dataset in which each camera pixel corresponds to an ODMR spectrum.
In this setting, conventional nonlinear fitting becomes prohibitively slow, often requiring tens of minutes to hours to analyze a single field of view containing $\gtrsim 10^5$ pixels.
This modality therefore provides an ideal testbed for our 1D-CNN inference framework, where one expects substantial gains in throughput while maintaining or improving robustness and accuracy.

Specifically, we apply the 1D-CNN to widefield magnetic imaging of a thin ($\sim 200$~nm) film of $\mathrm{Bi}_2\mathrm{Sr}_2\mathrm{CaCu}_2\mathrm{O}_{8+\delta}$ (BSCCO) mechanically exfoliated directly onto the diamond surface (Fig.~\ref{fig:B}a)~\cite{liu2025quantum}.
BSCCO is a layered, unconventional type-II superconductor with a transition temperature $T_c \approx 90$~K, and serves as a canonical cuprate platform for studying high-$T_c$ superconductivity and vortex matter.
In the superconducting state, the material expels magnetic flux, leading to a local suppression of the magnetic field (the Meissner effect).
During field-cooling, structural disorders or inhomogeneities within the sample lead to magnetic flux trapping, which manifests as the formation of isolated superconducting vortices.
Direct imaging of these vortices, and tracking their spatial organization and dynamics, provides direct access to pinning landscapes, dissipation mechanisms, and vortex-phase behavior in high-$T_c$ superconductors \cite{Hoffman2002Science_VortexCoresBi2212,Kim1996PRL_DecorationFieldCooledBi2212,Hoogenboom2000PRB_ShapeMotionVortexCoresBi2212}.
Accordingly, we map the magnetic stray field by performing wide-field ODMR measurements and extracting the local resonance splitting.

Figure~\ref{fig:B}b shows the magnetic field map reconstructed by the 1D-CNN, which clearly resolves a disordered lattice of individual vortices, i.e. a vortex glass state.
In particular, the CNN efficiently processes the megapixel array of ODMR spectra within $44$~s, $\sim 11$x faster than the hybrid method ($486$~s).
To quantify robustness, we evaluate the reconstruction error (RMSE of peak splitting, normalized by reference splitting) as a function of SNR (Fig.~\ref{fig:B}c), using again the longest-integration dataset as the ground truth.
We benchmark the standalone CNN against single-shot least-squares fitting under two initialization conditions.
Although fitting performs well when initialized near the optimum, its accuracy degrades rapidly at low SNR when the initial guess is slightly offset.
This sensitivity makes iterative fitting often unreliable for widefield imaging dataset, where a single global initialization cannot typically accommodate spatial variations across the field of view and thus leads to localized fitting failures (non-converging) and reconstruction artifacts.
In contrast, the CNN requires no initialization and maintains consistently low RMSE across the full SNR range.
These results confirm that our ML framework removes both the computational bottleneck and the initialization vulnerability that limit conventional large-scale quantum sensing image reconstruction.

\section{Conclusion and Outlook}

In summary, we demonstrate a robust deep-learning framework for the real-time analysis of NV center ODMR spectra.
Using a custom 1D-CNN trained on physically constrained synthetic data, our approach addresses key limitations of traditional nonlinear least-squares fitting.
In particular, it decouples parameter extraction from iterative optimization, providing nearly five orders of magnitude acceleration while avoiding the local minimum failures that can arise in gradient-based fitting.
We validate these advantages across diverse experimental settings, ranging from intracellular nanodiamond thermometry to robust widefield magnetic imaging of superconducting vortices.

Looking forward, our work opens several promising future directions.
First, the low computational footprint of the 1D-CNN makes it well suited for deployment on edge hardware such as field-programmable gate arrays (FPGAs).
Integrating inference directly into the acquisition pipeline could reduce latency to the microsecond scale, enabling closed-loop control and real-time adaptive sensing~\cite{krenn2023artificial,joas2021online}.

Second, the framework can be naturally extended to handle more complex ODMR spectral features, including hyperfine structure arising from nearby nuclear spins, as well as the simultaneous extraction of multiple physical parameters.
These include vector magnetic field, local temperature, the full stress tensor, and the relative orientation of the nanodiamond, obtained by combining information from different NV crystallographic groups within the ensemble~\cite{shim2022multiplexed,broadway2019microscopic,wang2025simultaneous}.
More broadly, although we focus here on NV centers, the same framework should be applicable to other quantum systems.
For example, recently emerging spin defects in two-dimensional hexagonal boron nitride often exhibit substantially broader and more intricate ODMR spectra because of their strong nuclear spin environment~\cite{gottscholl2021spin,stern2022room,he2025probing}.
In such systems, an ML-based approach could help resolve small splittings or shifts buried within strongly overlapping resonances, thereby improving measurement sensitivity.

Finally, the built-in uncertainty quantification provides a route toward ``smart" microscopy, in which the integration time is dynamically adjusted based on the predicted confidence of the real-time estimates.
We anticipate that ML-enabled analysis will become a standard component of the quantum-sensing toolbox, expanding the practical limits of speed and sensitivity in both biological and condensed matter applications.

\vspace{2mm}

\emph{Acknowledgements}: 
This work is supported by the Center for Quantum Leaps at Washington University, NSF NRT LinQ 2152221, NSF ExpandQISE 2328837 and NSF QIS 2514391.
S.M. acknowledges support from NIH grant R35GM142704.
C.Zhang acknowledges support from NSF 2503230.

\vspace{2mm}

\emph{Competing Interests}: J.R.B. is a member of the Scientific Advisory Board of LUCA Science, Inc.; receives research support from LUCA Science and Edgewise Therapeutics; is a consultant for Columbus Instruments, Inc.; has received royalties from Springer Nature Group and honoraria from Wiley; is an inventor on technology licensed to Columbus Instruments, Inc. with royalty rights; and is an inventor on pending patent applications related to the treatment of metabolic diseases (63/625,555), allergic diseases (US20210128689A1), and mitochondria transfer (018984/US).

\bibliographystyle{naturemag}
\bibliography{ref}

\end{document}


\title{Supplementary Information: A Deep-Learning-Boosted Framework for Quantum Sensing with Nitrogen-Vacancy Centers in Diamond}

\author{
Changyu~Yao,$^{1,*}$
Haochen~Shen,$^{1,*}$
Zhongyuan~Liu,$^{1}$
Ruotian~Gong,$^{1}$
Md Shakil~Bin~Kashem,$^{1}$
Stella~Varnum,$^{2}$
Liangyu Li,$^{3}$
Hangyue Li,$^{3}$
Yue Yu,$^{1}$
Yizhou~Wang,$^{1}$
Xiaoshui~Lin,$^{1}$
Jonathan Brestoff,$^{2,4}$
Chenyang Lu,$^{3,5}$
Shankar Mukherji,$^{1,4,6}$
Chuanwei Zhang,$^{1,4}$
Chong Zu$^{1,4}$
\\
\medskip
\normalsize{$^{1}$Department of Physics, Washington University, St. Louis, MO 63130, USA}\\
\normalsize{$^{2}$Department of Pathology and Immunology, Washington University School of Medicine, St. Louis, MO 63110, USA}\\
\normalsize{$^{3}$Computer Science \& Engineering, Washington University McKelvey School of Engineering, St. Louis, MO, USA, 63130}\\
\normalsize{$^{4}$Center for Quantum Leaps, Washington University, St. Louis, MO 63130, USA}\\
\normalsize{$^{5}$AI for Health Institute, Washington University, St. Louis, MO, USA., 63130}\\
\normalsize{$^{6}$Department of Cell Biology and Physiology, Washington University School of Medicine, St. Louis, MO 63110, USA}\\
}

\date{\today}

\maketitle

\tableofcontents

\section{Hamiltonian of the NV Center and ODMR Spectroscopy}

\subsection{Theory and Methods}

The ground-state Hamiltonian of the nitrogen-vacancy (NV) center can be described as the sum of contributions from the zero-field splitting (ZFS), the local electric and strain fields, and the external magnetic field \cite{doherty2013nv,maze2011properties}:
\begin{equation}
  H = H_\text{ZFS} + H_E + H_B \text{,}
\end{equation}
where the ZFS interaction is given by:
\begin{equation}
  H_\text{ZFS} = \mathbf{S} \cdot \mathbf{D} \cdot \mathbf{S} \approx D_\mathrm{gs} S_z^2 \text{,}
\end{equation}
the combined effect of the local electric field and transverse strain is expressed as:
\begin{equation}
  H_E = \Pi_z S_z^2 + \Pi_x(S_y^2 - S_x^2) + \Pi_y(S_x S_y + S_y S_x) \text{,}
\end{equation}
and the Zeeman interaction with the external magnetic field is:
\begin{equation}
  H_B = \gamma_e \mathbf{B} \cdot \mathbf{S} \text{.}
\end{equation}

In the presence of transverse fields and an axial magnetic field, the degenerate $|\pm 1\rangle$ spin states mix to form two new energy eigenstates, leading to the ODMR splitting measured experimentally\cite{rondin2014magnetometry,doherty2013nv}. The resulting energy splitting (measured by ODMR) between these hybridized states is given by
\begin{equation}
\label{eq:Splitting}
    \delta = 2\sqrt{\Pi_\perp^2 + (\gamma_e B_z)^2} \text{,}
\end{equation}
where $\Pi_\perp = \sqrt{\Pi_x^2 + \Pi_y^2}$ represents the effective transverse coupling, with the effect of $\Pi_z$ and $B_\perp$ negligible at small fields.

\begin{figure}[htbp!]
    \centering
    \includegraphics[width=0.98\linewidth]{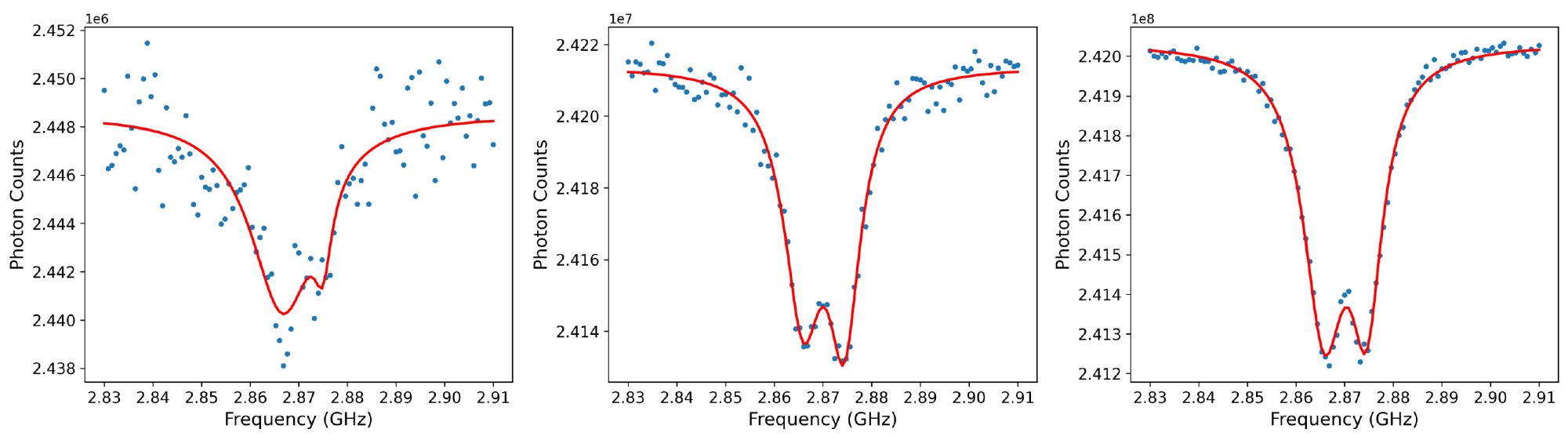}
    \caption{Experimental ODMR spectra obtained at zero field with SNR of 4.51, 13.54, and 42.75, respectively.}
    \label{fig:Raw}
\end{figure}

The ZFS tensor $\mathbf{D}$ is highly sensitive to variations in local strain and temperature, providing the fundamental basis for nanoscale sensing of these physical quantities \cite{acosta2010temperature,toyli2012thermometry}.
At zero external magnetic field, the spectrum shows an intrinsic splitting arising from the internal strain (Figure~\ref{fig:Raw}).
Therefore, we typically measure the center frequency at zero field for temperature sensing.

For external magnetic field measurements, we employ Eq.~\ref{eq:Splitting}.
In the low-to-intermediate field regime, however, the intrinsic transverse splitting $\Pi_\perp$ becomes comparable to the Zeeman splitting and cannot be neglected.
Therefore, standard measurement protocols involve initially acquiring a zero-field splitting map to serve as a baseline benchmark.
Then we can calculate the true $B_z$ from the obtained splitting map:
\begin{equation}
\label{eq:B}
    B_z
    = \sqrt{(\delta / 2 \gamma_e)^2 - (\Pi_\perp / \gamma_e) ^2}
    = \sqrt{(\delta / 2 \gamma_e)^2 - (\delta_0 / 2 \gamma_e) ^2}
\end{equation}

\subsection{Experimental Dataset and Data Quality}

In practice, the measured spectra may deviate from a perfect two-peak Lorentzian profile due to lattice inhomogeneity and spin noise. 
Nevertheless, we can still treat them as ideal Lorentzian line shapes while yielding sufficiently accurate parameter estimations \cite{poole1996electron}.

The spectral contrast is primarily governed by polarization and microwave delivery, which typically remain fixed.
However, extending the collection time increases the total photon counts, thereby reducing relative noise and enhancing the signal-to-noise ratio (SNR).
Figure~\ref{fig:Raw} presents the experimental spectra with SNRs of 4.51, 13.54, and 42.75, respectively.

\subsection{Extented Data of Superconducting Vortices}

\begin{figure}[htbp!]
    \centering
    \includegraphics[width=0.98\linewidth]{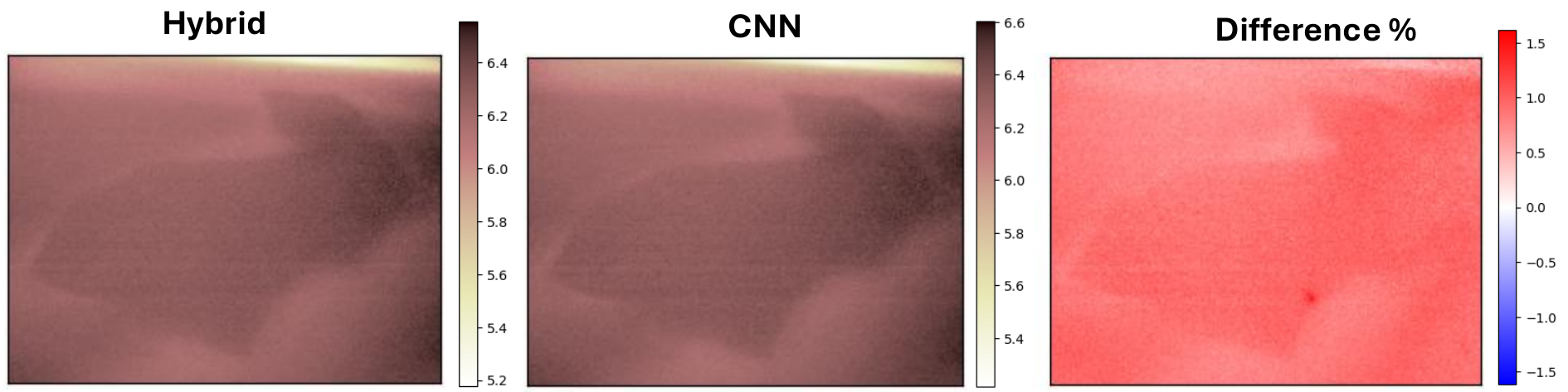}
    \caption{Intrinsic splitting map ($\Pi_\perp / \gamma_e$) measured without an external field.}
    \label{fig:IntrinsicSplitting}
\end{figure}

\begin{figure}[htbp!]
    \centering
    \includegraphics[width=0.98\linewidth]{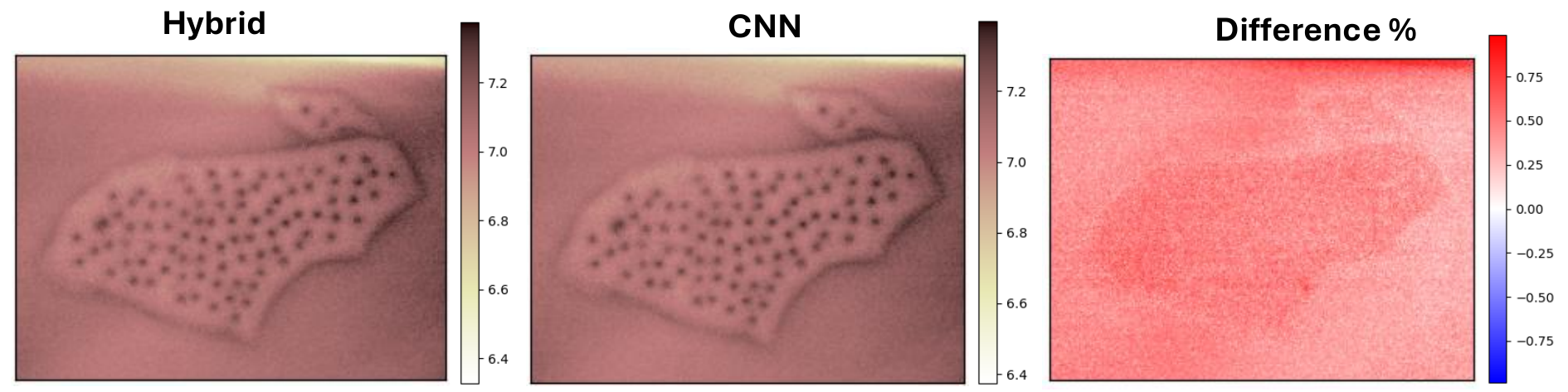}
    \caption{Splitting map ($\delta / 2 \gamma_e$) under the applied external field.}
    \label{fig:FieldSplitting}
\end{figure}

\begin{figure}[htbp!]
    \centering
    \includegraphics[width=0.98\linewidth]{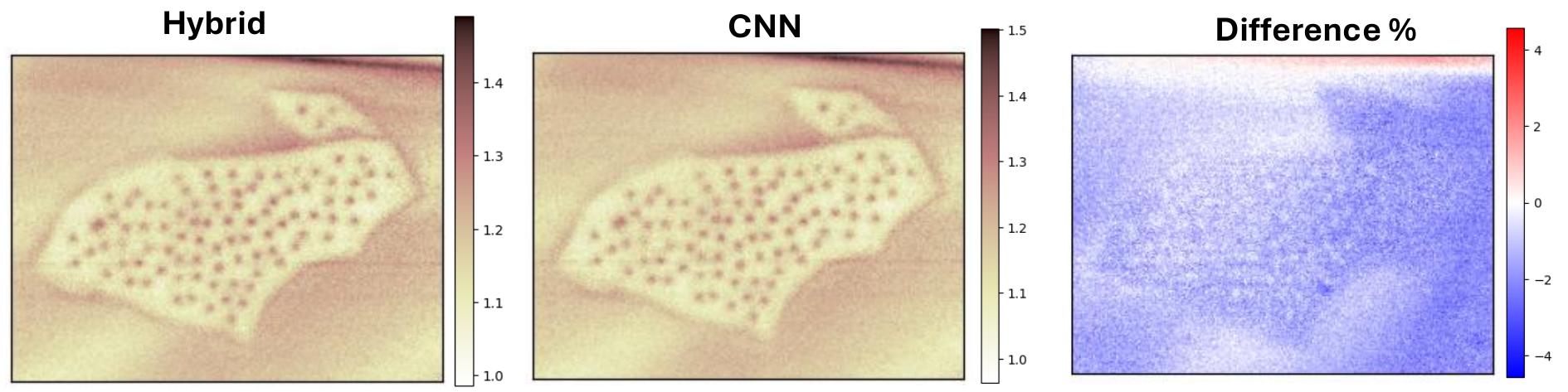}
    \caption{Magnetic field map ($B_z$) calculated from Eq.~\ref{eq:B}}
    \label{fig:BMap}
\end{figure}

We follow the standard protocol on measuring the superconducting vortices, in which the applied magnetic field is in the low-field regime and comparable with the intrinsic splitting.
To accurately evaluate the local magnetic field, we first measured the intrinsic splitting at zero external magnetic field to map the spatial distribution of $\Pi_\perp$ (Fig.~\ref{fig:IntrinsicSplitting}).
Following this baseline characterization, we measured the total energy splitting under the applied field (Fig.~\ref{fig:FieldSplitting}).
Following equation~\ref{eq:B}, we isolated the Zeeman contribution and calculated the pure out-of-plane magnetic field distribution $B_z$. This procedure yielded the final magnetic field map of the superconducting vortices presented in Fig.~\ref{fig:BMap}.

\section{Model Details and Model Studies}

In this section, we provide a detailed description of our model.
This study focuses only on the extraction of two-peak parameters—a task central to temperature sensing, weak magnetic field detection, and single-group NV measurements.
We outline the specific configurations used for experimental data analysis and the generation of synthetic training data. Additionally, we present the scaling laws governing our model's hyperparameters.

\subsection{Model Details}

\begin{figure}
    \centering
    \includegraphics[width=0.98\linewidth]{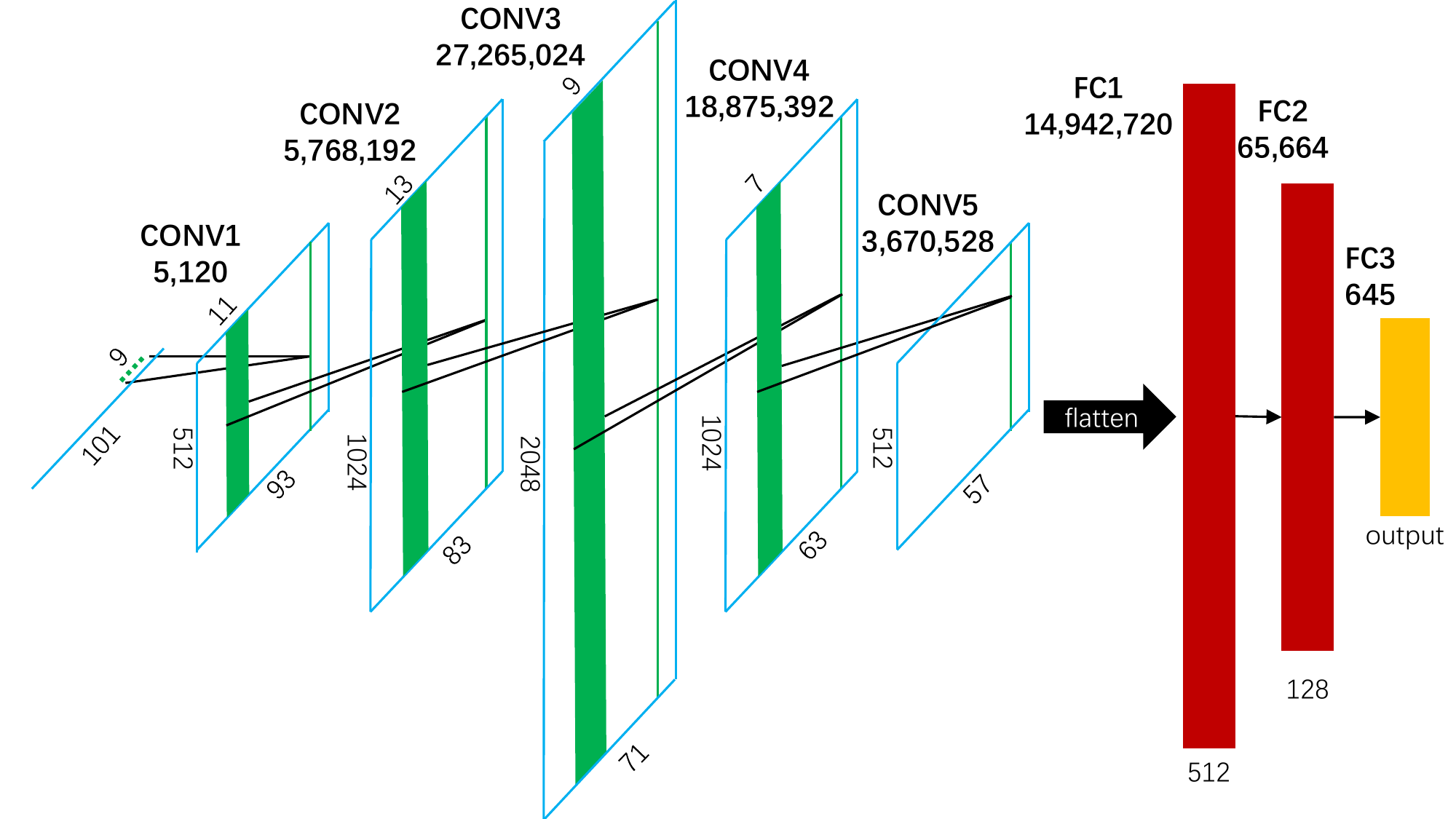}
    \caption{The model archtecture and parameter calculation.}
    \label{fig:Net}
\end{figure}

Here we show our model architecture in Figure~\ref{fig:Net}: a 1D-CNN designed to process spectra of length $101$, following the general convolutional neural network framework widely used for extracting localized features from structured data \cite{lecun1998gradient,krizhevsky2012imagenet}. To preserve the precise spatial relationships without introducing edge artifacts, no padding is applied to any of the convolutional layers. Consequently, the sequence length at each layer strictly follows the formula: Output Length = Input Length - Kernel Size + 1. All hidden layers utilize the Rectified Linear Unit (ReLU) activation function, and no dropout is applied during the training process.
The network consists of five consecutive 1D convolutional layers followed by three fully connected (FC) layers.
The total number of trainable parameters in the network is \textbf{70,593,285}.

\subsection{Data Synthesis}

We use a perfect two-peak Lorentzian function as the training input:
\begin{equation}
    S(x, x_1, x_2, w_1, w_2, C_1, C_2) = 1 - C_1 L(x, x_1, w_1) - C_2 L(x, x_2, w_2)\text{,}
\end{equation}
where $L$ represents the Lorentzian spectrum:
\begin{equation}
    L(x, x_0, w) = \frac{1}{1 + \frac{(x-x_0)^2}{w^2}}
\end{equation}
and $x_1, x_2$ denote the dip positions, $w_1, w_2$ the dip widths, and $C_1, C_2$ the dip contrasts.

To match the model's input dimensions, the variable $x$ consists of $101$ uniformly spaced points between $0$ and $1$, representing the normalized frequency.

To generate a diverse set of spectra, we randomize the parameters. For the widths ($w_1$ and $w_2$), we first sample a base width $w$ from a uniform distribution between $0.02$ to $0.09$. To account for spectral asymmetry, we introduce an asymmetry parameter $a_{\text{as}}$, sampled from the positive half of a normal distribution ($\sigma = 0.083$) and clamped at a maximum of $0.25$. With a $50\%$ probability, we set $w_1 = w$ and $w_2 = w(1 - a_{\text{as}})$; otherwise, $w_1 = w(1 - a_{\text{as}})$ and $w_2 = w$. The contrasts $C_1$ and $C_2$ are generated using a similar approach, but with a base contrast $C$ drawn uniformly from $0.012$ to $0.15$.

The dip positions $x_1$ and $x_2$ are determined using a center-splitting model. The center $c$ is sampled uniformly from $0.35$ to $0.65$, and the splitting distance $s$ is sampled uniformly from $0.02$ to $0.2$. To prevent the dips from merging, $s$ is clamped to a minimum value of $0.75(w_1 + w_2)/2$. The positions are then calculated as $x_1 = c - s/2$ and $x_2 = c + s/2$. This ensures that $x_1$ consistently represents the left dip and that the two dips maintain a minimum separation, which physically accounts for intrinsic splitting.

Following the generation of the ideal Lorentzian profiles, we introduce photon shot noise. We simulate a total photon count drawn uniformly between $180,000$ and $3,600,000$ and apply point-wise Poisson noise to mimic realistic experimental conditions.

Finally, we apply Z-score normalization by subtracting the mean and dividing by the standard deviation. As a result, the overall amplitude of the data is standardized, regardless of the absolute dip contrast. However, a consequence of this normalization is that a spectrum with contrasts $(C_1, C_2)$ becomes indistinguishable from one with $(1, C_2/C_1)$, which renders the model incapable of directly extracting absolute contrast values from the processed signal.

\subsection{Hyper-parameter Optimization Studies}

\begin{figure}[htbp!]
    \centering
    \begin{minipage}[t]{0.48\linewidth}
        \centering
        \includegraphics[width=\linewidth]{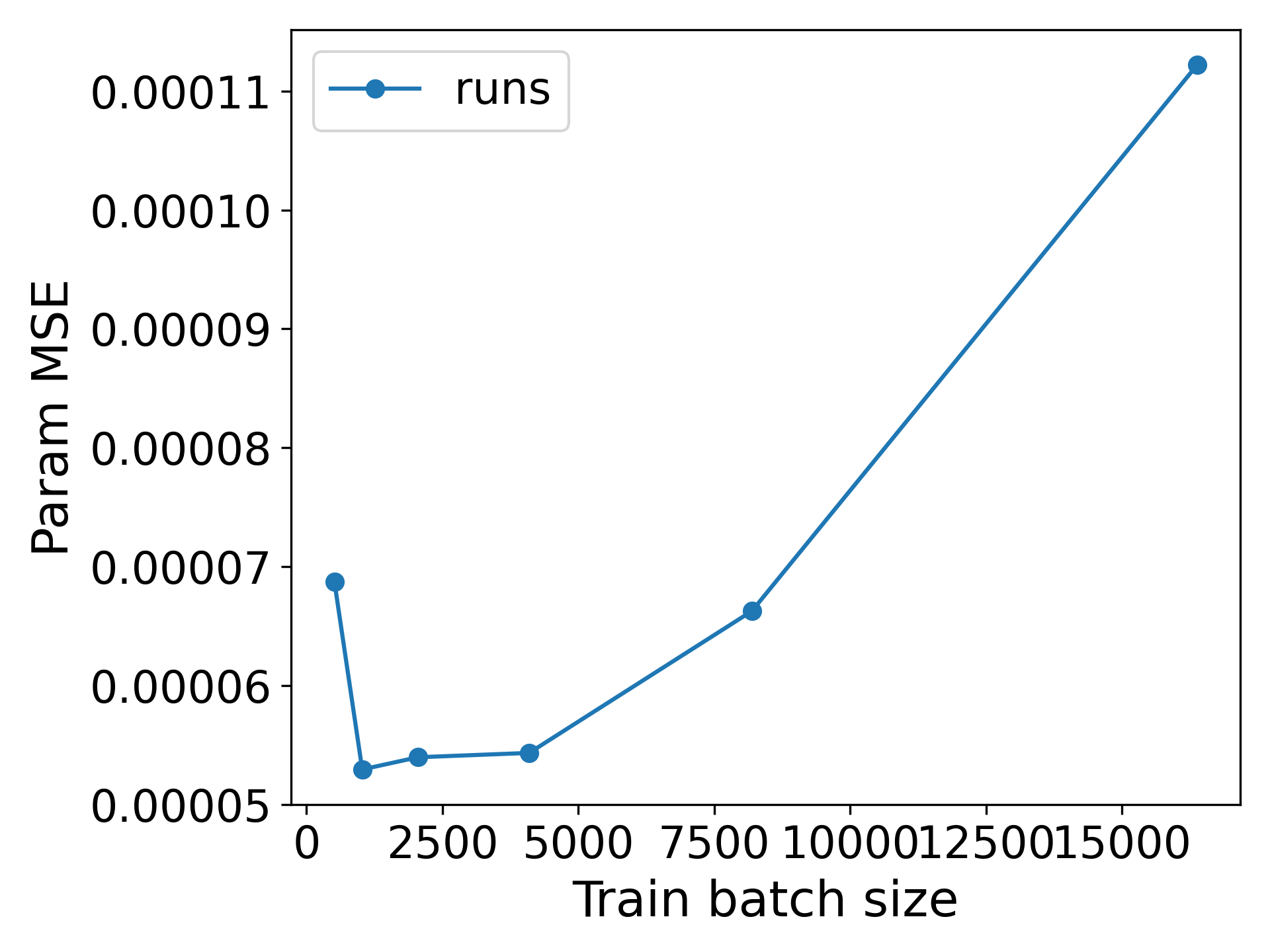}
    \end{minipage}
    \hfill
    \begin{minipage}[t]{0.48\linewidth}
        \centering
        \includegraphics[width=\linewidth]{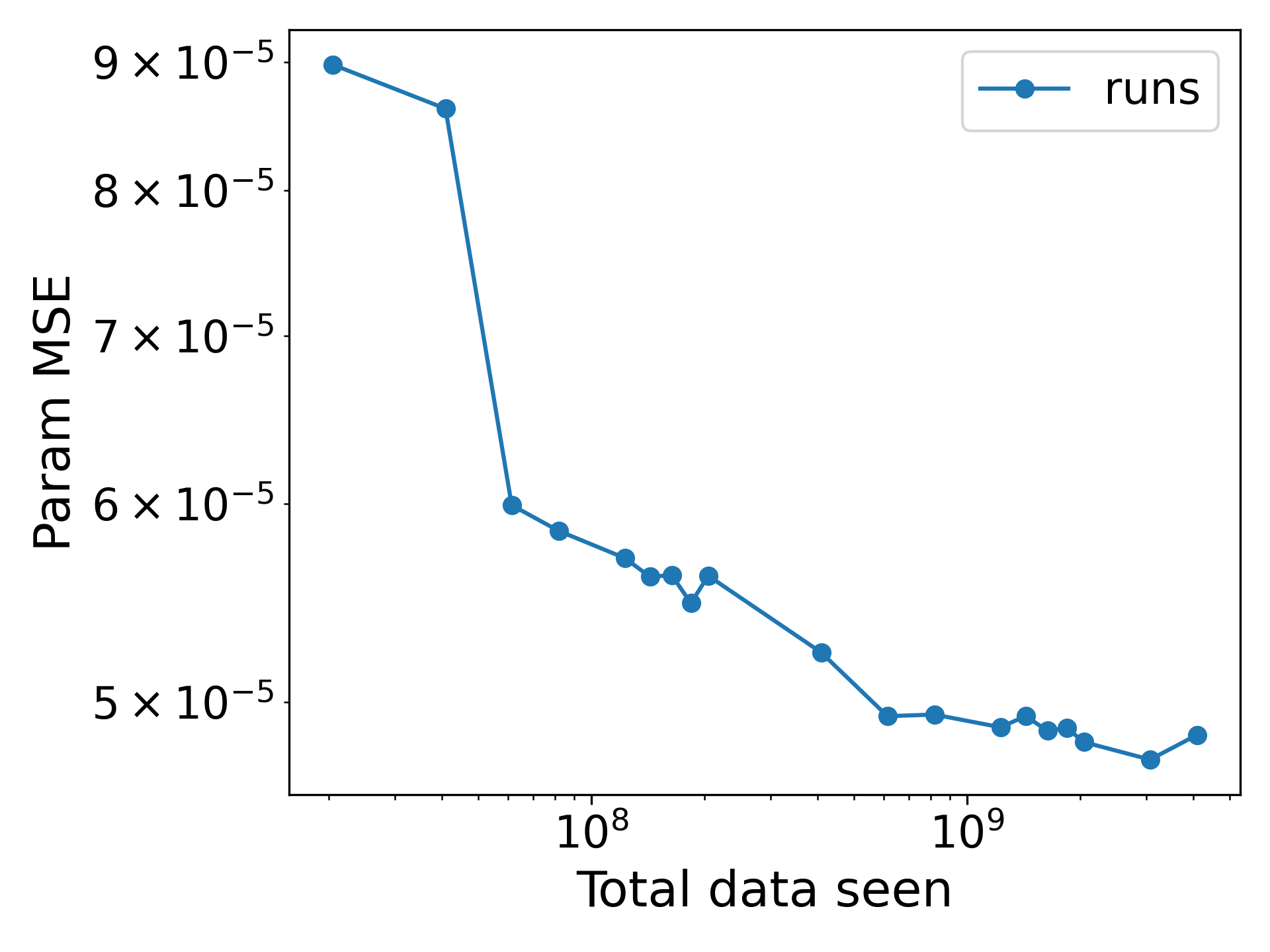}
    \end{minipage}

    \caption{Scaling behavior of the model. The left figure is the batch size scaling result, showing an optimum around 2048-4096; The right figure shows the data scaling of the model $[128,256,512,512]$, showing a plateau at around $3\times10^9$.}
    \label{fig:scaling_combined}
\end{figure}

\begin{figure}[htbp!]
    \centering
    \includegraphics[width=0.9\linewidth]{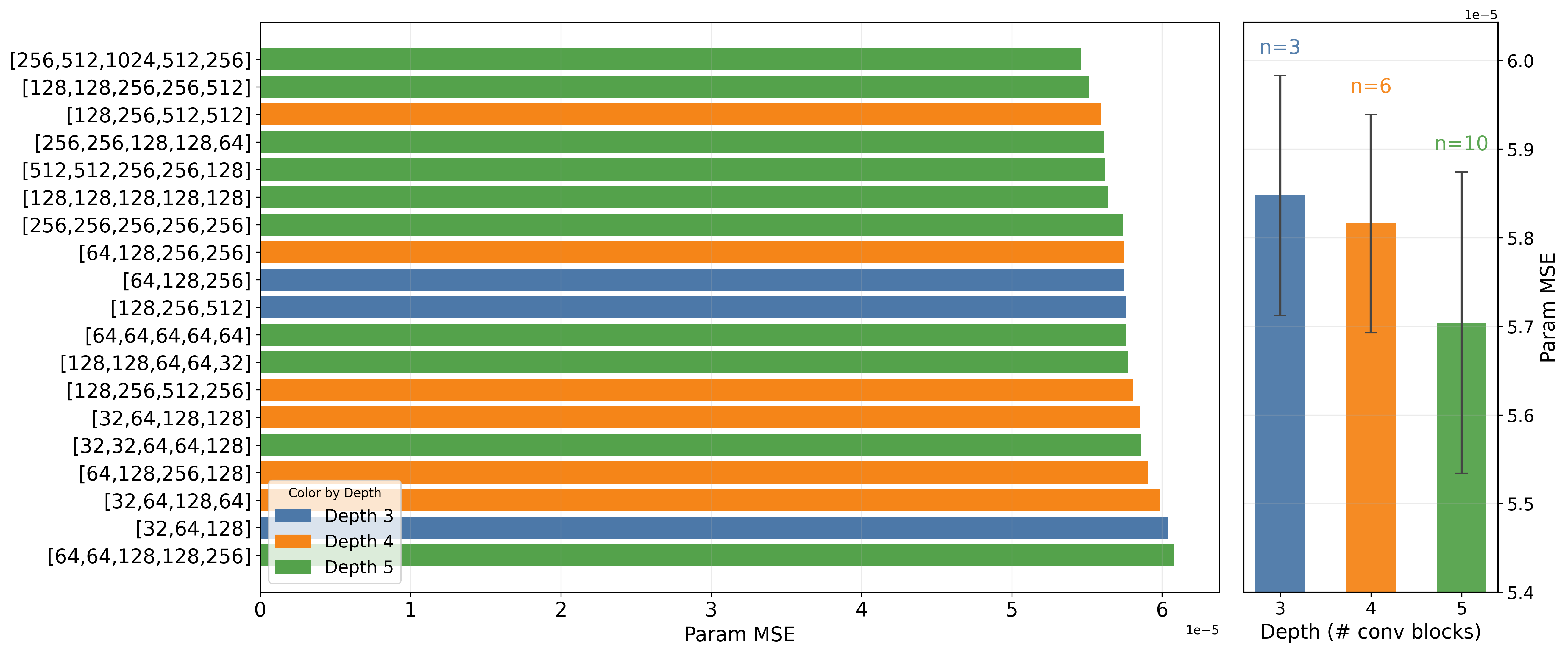}
    \caption{Performance of 19 models at $2.048\times10^9$ data seen (left) and performance categorized by different depth (right).}
    \label{fig:all_and_depth}
\end{figure}

\begin{figure}[htbp!]
    \centering
    \includegraphics[width=0.6\linewidth]{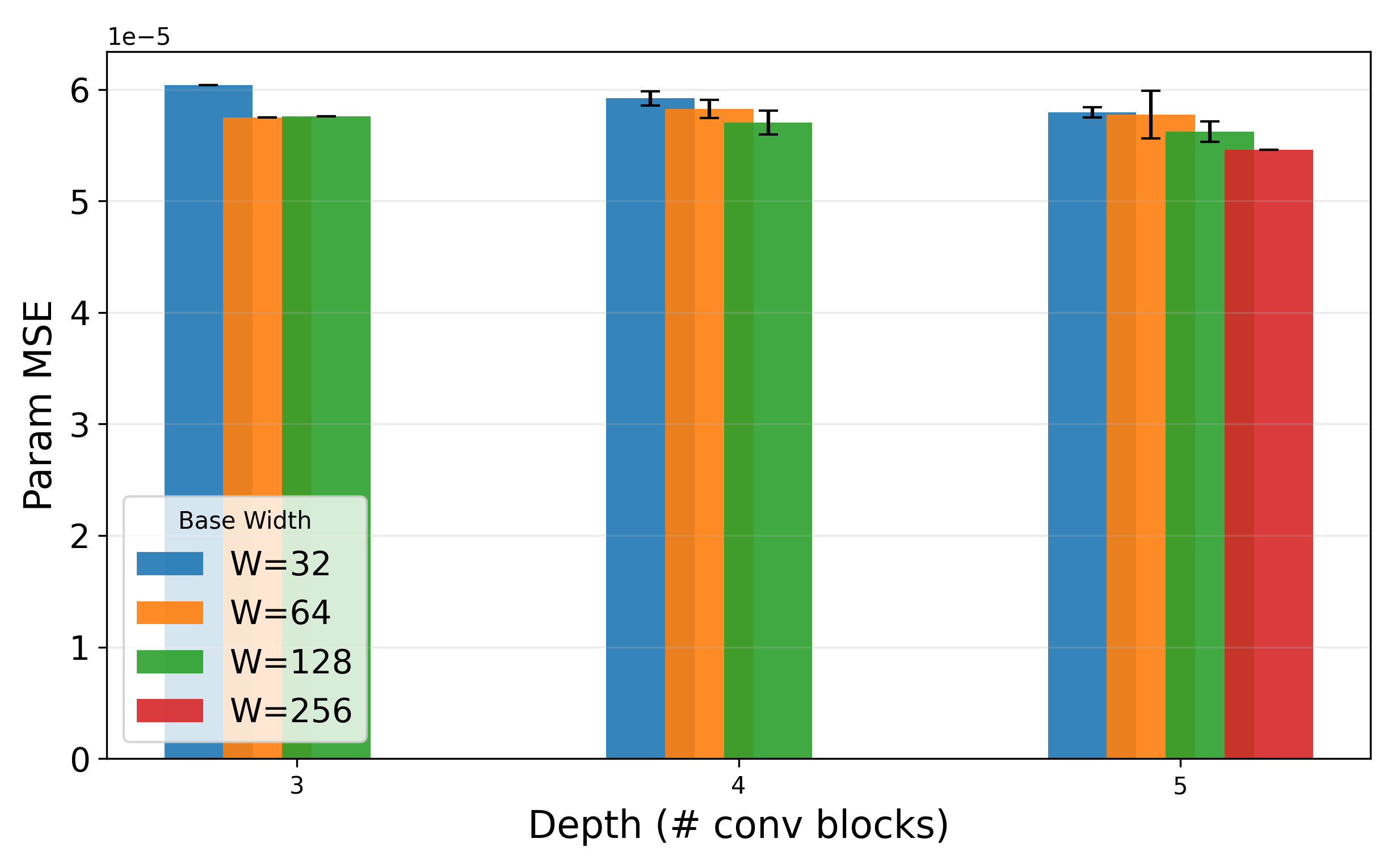}
    \caption{Performance categorized by width among each depth. The base width is the smallest number of channels across each block.}
    \label{fig:width}
\end{figure}

In this section, we evaluate how model hyperparameters and training configurations influence final performance. This section is organized following the analysis of three aspects of scaling law: \textbf{batch size scaling} (Figure~\ref{fig:scaling_combined} left), \textbf{data scaling} (Figure~\ref{fig:scaling_combined} right) and \textbf{parameter scaling} (Figure~\ref{fig:all_and_depth}, \ref{fig:width}). We denote specific model configurations as $[N_i]$, where $N_i$ represents the number of features (channels) in the $i$-th convolutional layer.

A common optimizer and learning rate scheduler is shared by all experiments:
\begin{itemize}
    \item 5\% warmup that starts at learning rate $\eta = 0.1\eta_{\mathrm{max}}$
    \item Cosine annealing scheduler after the warmup that ends at last step with $\eta_{\mathrm{min}}=1\times10^{-5}$ and $\eta_{\mathrm{max}}=1.5\times10^{-3}$
    \item AdamW optimizer with $(\beta_1,\beta_2)=(0.9,0.95)$ and weight decay $\lambda=1\times10^{-2}$ \cite{loshchilov2019decoupled}. Other hyperparametes kept in PyTorch default.
    \item Batch size $B=2048$ except batch scaling experiment.
\end{itemize}

The training data is generated on-the-fly by the double-Lorentzian model, enabling convenient scaling-law studies in a controlled way and also preventing overfitting.

The batch size $B$ is a critical training hyperparameter that directly governs optimization dynamics. Increasing $B$ typically reduces stochastic gradient noise and enables a proportionally larger learning rate according to the linear scaling rule for deep neural network training \cite{goyal2017accurate}. This often accelerates training convergence in terms of wall-clock time. However, batch size cannot be increased indefinitely due to hardware memory constraints and, more critically, the point of diminishing returns in data efficiency.
As shown in Figure~\ref{fig:scaling_combined} (left), we observe an optimal batch size range between $2048$ and $4096$ when employing a linear scaling rule for the learning rate, defined as $\eta = 0.0015 \cdot B/2048$. Beyond this threshold, the model requires significantly more total samples to achieve the same level of convergence, indicating a drop in sample efficiency that outweighs the benefits of larger optimization steps.

Data scaling determines the volume of training samples required for a model to adequately generalize. In our context, where the synthetic data generator provides an effectively infinite stream of samples, the primary concern is the total cumulative data needed to fully characterize the underlying double Lorentzian model.
Figure~\ref{fig:scaling_combined} (right) illustrates the relationship between cumulative training samples and Mean Squared Error (MSE) for a representative architecture, $[128, 256, 512, 512]$. The performance reaches a plateau after approximately $3 \times 10^9$ samples. While the saturation point may vary slightly with model capacity, we observe that it remains relatively consistent across architectures within the same family. In practice, we utilize a slightly larger dataset to ensure that models reach a state of full optimization regardless of minor variations in parameter count.


To evaluate the impact of model capacity on performance, we conducted an architectural search across 19 distinct configurations, assessing each at a fixed training milestone of $2.048 \times 10^9$ samples. While Figure~\ref{fig:all_and_depth} left suggests a general trend of performance gains as model depth increases, a more granular analysis reveals that the $1\sigma$ confidence intervals for different depths frequently overlap (Figure~\ref{fig:all_and_depth} right). This indicates that, within the tested range, increasing the number of layers does not drastically improve the accuracy. Furthermore, deeper architectures introduce a higher number of trainable parameters, which typically necessitates larger datasets to reach peak optimization.
In contrast, increasing the base width—defined as the minimum number of channels across all blocks—results in a more consistent performance enhancement across all tested depths, even when accounting for error margins (Figure~\ref{fig:width}). Despite these trends, the overall sensitivity of the MSE to changes in parameter count and depth remains relatively low. 
Overall, these results suggest a diminishing return on architectural complexity for the double Lorentzian model, allowing for a practical compromise between computational efficiency and predictive accuracy.

\newpage

\bibliographystyle{ieeetr}
\bibliography{ref.bib}